# Controlled Growth of Large-Size 2D Selenium Nanosheet and Its Electronic and Optoelectronic Applications

Jing-Kai Qin[1,3,†], Gang Qiu[1,†], Jie Jian[2], Hong Zhou[1], Ling-Ming Yang[1], Adam R. Charnas[1], Dmitry Zemlyanov[1], Cheng-Yan Xu[3], Xian-Fan Xu[4], Wen-Zhuo Wu[5], Hai-Yan Wang[2,1], Peide D. Ye[1,*]

[†] These authors contributed equally to this work.

[1.] School of Electrical and Computer Engineering, Purdue University, West Lafayette, IN 47907, USA

[2.] School of Materials Science and Engineering, Purdue University, West Lafayette, In 47907, USA

[3.] School of Materials Science and Engineering, Harbin Institute of Technology, Harbin 150001, Peoples' Republic of China.

[4.] School of Mechanical Engineering, Purdue University, West Lafayette, In 47907, USA

[5.] School of Industry Engineering, Purdue University, West Lafayette, In 47907, USA

* Address correspondence to: yep@purdue.edu

**ABSTRACT**

Selenium has attracted intensive attention as a promising material candidate for future optoelectronic applications. However, selenium has a strong tendency to grow into nanowire forms due to its anisotropic atomic structure, which has largely hindered the exploration of its potential applications. In this work, using a physical vapor deposition method, we have demonstrated the synthesis of large-size, high-quality 2D selenium nanosheets, and the minimum thickness of which could be as thin as 5 nm. The Se nanosheet exhibits a strong in-plane anisotropic property, which is determined by angle-resolved Raman spectroscopy. Back-gating field-effect transistors (FETs) based on Se nanosheet exhibit p-type transport behaviors with on-state current density around 20 mA/mm at $V_{ds}$=3V. Four-terminal field-effect devices are also fabricated to evaluate the intrinsic hole mobility of selenium nanosheet, and the value is determined to be 0.26 cm$^2$ V$^{-1}$ s$^{-1}$ at 300 K. The selenium nanosheet phototransistors show an excellent photoresponsivity up to 263 A/W, with the rise time of 0.1s and fall time of 0.12s. These results suggest that crystal selenium as a 2D form of 1D van der Waals solid, opens more feasibility to explore device applications.

**Keywords:   selenium; 1D crystal structure; 2D nanosheet; electrical transport; photoresponse**

Trigonal selenium (*t*-Se) is a one-dimensional elemental semiconductor material, in which Se atoms are covalently connected in a spiral chain along c axis with two adjacent atoms rotated by 120°. As shown in **Figure 1a and b**, all these atomic chains are stacked together in their radical direction by weak van der Waals interactions to form a hexagonal lattice structure. The *t*-Se has attracted extensive attention due to its interesting properties such as high photoconductivity,[1-5] high piezoelectricity,[6-8] thermoelectricity[9] and nonlinear optical responses.[10, 11] The high photoconductivity makes it a promising candidate for applications in high-efficiency solar cells and optoelectronic devices.[12, 13] Recent reports have shown that Se nanosheet photodetectors exhibit a high responsivity of 100 mA/W at 620 nm light illumination with ultrashort rise/decay time (1.4/7.8 ms).[14]

A variety of Se-based nanostructures have been obtained by different synthesis methods, such as hydrothermal solution process and microwave-assisted synthesis in ionic liquids.[14, 15] Compared with solution based techniques, which always introduce impurities due to the complex chemical processes involved, a vapor phase approach should be more efficient to obtain high quality products.[16, 17] Due to the anisotropic chain-like crystal structure, Se tends to form 1D-structures, such as nanowires, nanotubes and nanobelts. These 1D structures with high ratio of edges to bulk always exhibit poor electrical transport behaviors due to localized states and contacts, and the issue of electrical noise becomes critical with decreasing size as described by Hooge's rule.[18, 19] To expand its applications in high-performance electronic and optoelectronic devices, it would be useful to expand Se nanostructures to a two-dimensional form.

Recently, tellurium (Te), which possesses the same crystal structure as Se, has been successfully synthesized in a 2D form using a substrate-free solution process, and 2D tellurium (termed as tellurene) flakes exhibit strong in-plane anisotropic properties and high carrier mobility.[20-22] Inspired by this success, using a physical vapor deposition (PVD) method, we successfully obtained highly crystalline Se nanosheets with interesting zigzag edge structure, with a minimum thickness of Se 5nm. Scanning transmission electron microscopy (STEM) and angle-resolved Raman spectroscopy confirm that the single-crystal Se nanosheet has an oriented growth direction along $<1\bar{2}10>$, which is very different from previous studies.[16, 17] To the best of our knowledge, no works have been reported about the synthesis of such large-size 2D ultrathin Se nanosheets. Back-gated field-effect transistors (FETs) based on our 2D PVD-grown Se nanosheet exhibit p-type transport behaviors with on-state current density around 20 mA/mm at $V_{ds}$=3V. Four-terminal field-effect devices are also fabricated to evaluate the intrinsic hole mobility of Se nanosheet, and the value is determined to be 0.26 cm$^2$ V$^{-1}$ s$^{-1}$ at 300K. Se nanosheet phototransistors are found to have an excellent photoresponsivity (263 A/W) with a 0.10 s rise time and a 0.12 s fall time. This work demonstrates that crystal selenium nanosheet, which is built up by 1D van der Waals material selenium, has very interesting properties for further exploration of device applications as a type of 2D semiconductor material.

**Results and Discussion**

The Se nanosheets were obtained using PVD method, and the typical growth setup used is illustrated in **Figure 1c**. Se powder was placed in a quartz glass tube as the precursor. A plan-view SEM image of Se nanosheets as-grown on a Si substrate is shown in **Figure 1d**, in which inclined nanosheets (as indicated by arrows) with density over 50% were uniformly distributed on the surface of polycrystalline Se films. **Figure 1e** shows the representative morphology of the Se nanosheets at high magnification in SEM. Most of the nanosheets exhibit a saw-like structure with zigzag edges on single narrow side. The average width of nanosheets is about 8 $\mu$m, and the maximum length of the nanosheets could reach up to 50 $\mu$m. In SEM images, the nanosheets are almost electron transparent, suggesting that they are very thin. It is worth noting that about 10% of the Se nanosheets have bi-lateral zigzag edges, exhibiting a feather-like structure (**Figure 1f**). The growth mechanism for these 2D Se nanosheets will be discussed in the following section.

To further indentify the surface morphology, we transferred the nanosheets onto a Si/SiO$_2$ substrate using the scotch tape method (the typical transfer procedure is shown in **Figure S1**). **Figure 2a** shows the optical microscopy image of Se nanosheets. Consistent with the SEM images, most of the Se nanosheets exhibit an irregular quadrangular shape with zigzag edges on unilateral side, and the rest of them have twin structures with mirror symmetry. Atomic force microscopy (AFM) was employed to determine the thickness and topography of the nanosheets. As the 3D view shown in **Figure 2b**, the minimum measured thickness of Se nanosheets is 5 nm, and the surface is atomically flat with a surface roughness ($R_a$) of 190 pm, even less than that of the SiO$_2$ surface (about 200 pm), indicating that the Se nanosheets are highly crystallized with good quality. **Figure 2c and d** show the height profiles of the saw-like and twin structure selenium nanosheets, with an average thickness of 15 nm. It should be noted that the twin structure could be clearly detected according to contrast difference by optical microscopy, while the corresponding AFM image shows a uniform height profile. This result suggests that the bi-crystalline Se nanosheet divided from the middle has two separate single crystal grains with different orientations. XPS analysis was also conducted to determine the quality of Se nanosheets (**Figure S2**).The strong peaks located at 55.5 and 54.3 eV correspond to Se $3d_{3/2}$ and $3d_{5/2}$ binding energy of $Se^{(0)}$ respectively. Se $3d$ oxidized peak located 59.9 eV cannot be detected, indicating that the Se nanosheets are elemental crystals without obvious oxidation.[23, 24]

Transmission electron microscopy (TEM) was employed to identify the microstructure and growth directions of the Se nanosheets. The nanosheets could be directly transferred onto Cu grids. The low-magnification bright-field TEM image (**Figure 3a**) shows a 2D saw-like structure, similar to what was observed in the optical microscopy image. High-angle annular dark field STEM (HAADF-STEM) image could provide Z-contrast with atomic lateral resolution (Z =atomic number). As shown in **Figure 3b**, Se helical atomic chains could be clearly resolved with fringe spacing of (0001) lattice planes about 5.0 Å. The corresponding selected area electron diffraction (SAED) image, which was obtained along the [10$\bar{1}$0] zone axis of an individual nanosheet, exhibits a set of two-fold and rotational symmetry pattern, indicating the Se nanosheet is highly crystalline. Combined with the HAADF-STEM image and SAED pattern, we can determine that the Se nanosheets have first grown along the $<0001>$ direction into nanoribbons, and then these nanoribbons expand in parallel along $<1\bar{2}10>$ direction, leading to the formation of large-area nanosheet with zigzag edges. TEM analysis was also conducted to determine the crystal structure of the feather-like Se nanosheets with twin structure. As the bright-field TEM images with different magnification shown (**Figure 3e and f**), the angle of two twinned grains is about 124°. By carefully correlating the SAED pattern obtained from the boundary area (**Figure 3g**), it could be found that $\{0\bar{1}04\}$ diffraction spots are perpendicular to the twin boundary, indicating that the twin boundary is a $\{0\bar{1}04\}$ twin. EDX and EELS are also confirmed that the synthesized nanosheet is elemental Se crystal (**Figure S3**).

The in-plane anisotropy of Se nanosheet is investigated by angle-resolved Raman spectroscopy at room temperature. The angle between laser polarization direction and [0001] helical chain direction of the nanosheet is defined as $\theta$, which could be tuned by rotating the sample in a step of 15° during measurement. **Figure 4a** shows the typical Raman spectrum of 15nm thick nanosheet with the angle $\theta$ of 45°. Being consistent with the previous observations in bulk selenium,[25, 26] three active Raman photon modes are clearly observed. Raman peaks located at ~233 and 237 cm$^{-1}$ are related to the $E_2$ and $A_1$ modes respectively. Meanwhile, one degenerate $E_1$ modes caused by *a*-axis rotation is also indentified. **Figure 4b** depicts the evolution of the Raman spectrum as the sample is rotated in the step of 30° from −90° to 90°, and clear intensity change could be observed. The peak intensity of different modes are extracted by fitting with Gauss function and plotted into the corresponding polar figures (**Figure 4c-d**). It should be noted that the degenerate $E_1$ mode is hard to be extracted due to the relative weak intensity. Both the $E_2$ and $A_1$ modes exhibit significant intensity change with the polarization angle. The $A_1$ mode is a maximum at an angle of ~90°, which corresponds to the direction vertical to the Se chains, while mode $E_2$ has the maximum value at an angle of ~45°. Our results suggest that the in-plane anisotropic properties of Se nanosheets could be easily identified by Raman spectroscopy. It could be also confirmed that the Se nanosheets are grown along $<1\bar{2}10>$ direction, which matches well the STEM results.

Trigonal selenium possesses a highly anisotropic crystal structure. Under thermodynamic equilibrium condition, it tends to grow into 1D structures along the [0001] direction. However, the products in our experiments exhibit 2D structure with very small thickness, which is very different from the previous bulk Se. Time-dependent experiments were conducted to explore the growth mechanism of Se nanosheets. **Figure 5a** shows the typical morphology of products when the growth temperature just reaches 160°C. Amorphous nanospheres with diameter ranging from 10-30 μm appeared on Si substrate surface. With temperature rising up to 210°C, amorphous nanospheres start to crystallize and form polycrystalline films, with numerous short nanorods protruding out of the surface (**Figure 5b and Figure S4**). Based on previous studies, heat treatment could accelerate the transformation of amorphous selenium to crystalline selenium,[27] and Imura *et al*. has successfully prepared polycrystalline films consisting of nanoparticles by *in situ* thermal treatment for photodetector application.[28] The preferable growth of 1D Se nanorods is kinetically favored, since the binding energy of Se atoms along c-axis is much higher than that along $<1\bar{2}10>$ directions.[16] Besides, these 1D nanorods possess high chemical activity especially at the vertexes and ridges, which could play the role of roots for the subsequent nanosheets growth.[29] Interestingly, as the reaction continues, the growth along c-axis is partly hindered, and the nanorods tend to expand in parallel along $<1\bar{2}10>$ direction to form 2D thin nanosheets as shown in **Figure 5c**. Under the same magnitude of super saturation, the growth of Se nanostructure is mainly driven by surface energy as well as effective activation sites count.[30] In the initial crystallization, 1D growth of nanorod is dominant due to the distinct surface energy difference. However, after the nanorod has grown to certain length, $Se_2$ (g) molecules would seldom fall within the end cavity of a nanorod due to the large free path at such high temperature.[30] It is more likely that they would impinge on nanorod surface and diffuse from site to site until they encounter defect or void for crystalline, thus the growth along $<1\bar{2}10>$ direction would be enhanced. Actually, at the lateral side of Si substrate, where the downstream $Se_2$(g) molecules have greater odds of encountering (0001) facet of nanorod, the nanorod would continue to grow along $<0001>$ direction into nanowire with length up to 200 $\mu$m (**Figure S5**). **Figure 5d** illustrates the products' morphology when the growth duration is up to 60 min, the nanorods have completely turn into long, saw-like ultrathin nanosheets.

The formation of Se nanosheets could be well explained by the vapor-solid (VS) growth mechanism,[30, 31] and similar growth phenomena have been reported in the preparation of ZnSe nanobelts.[32] It is worthy being noted that some of the nanosheets exhibit feather-like structure with mirror twins. We believe that they were developed from sections of intersected nanorods, which are introduced at initial crystalline stage (**Figure S6**). These crossed nanorods could act as growth roots, and enable the identical lateral growth on both sides of the nanosheet, leading to the symmetrical structure.[33, 34]

Se nanosheet FETs were fabricated using electron beam lithography (EBL), thermal evaporation, and lift-off process. Ni/Au (30/100nm) was selected as metal contacts, which could significantly reduce the contact resistance in p-type FET devices due to the relatively high work function.[35] **Figure 6** shows the electrical characteristics of a typical Se nanosheet FET device with channel thickness of 16 nm. It shows typical p-type transport behavior with a high current on/off ratio over $10^6$. The presence of hydrogen and hydroxyl terminations on the nanosheets surface is considered to be the main reason for the p-type conduction of selenium FET, which has been demonstrated in Se nanowire and nanobelts.[17,36] The Se nanosheet FET presents a relative low on-state current, with the maximum value around 20 mA/mm at $V_{ds}$=3V. Four-terminal field-effect devices are fabricated to evaluate the intrinsic hole mobility of Se nanosheets, as shown in **Figure S7**. The mobility could be extracted from the *G versus* $V_{bg}$ curves in the $-100 < V_{bg} < -30$ V range using the expression $\mu = L_{in}/W \times (1/C_{bg}) \times dG/dV_{bg}$, and the value of hole mobility is expected to be 0.26 cm$^2$V$^{-1}$s$^{-1}$ at 300 K. Compared with other p-type semiconductor materials such as tellurene and black phosphorus,[21, 37-39] Se nanosheet has a low hole mobility similar to those values reported in literature on bulk selenium or selenium nanobelts.[40-42]

Because of its low mobility and special band-structure, it has the highest Seebeck coefficient (+1250 µVK$^{-1}$) among all elements and has the potential for thermoelectric applications. It is worth noting that the Se nanosheets and the transistors exhibit a good stability in ambient condition (**Figure S8**). Even after 15 days' exposure to air, the surface morphology of Se nanosheet almost does not change, and no significant degradation of device performance.

Selenium is known as an excellent material candidate for high sensitivity optoelectronics. Herein, we also examine the optoelectronic performance of Se nanosheet phototransistors, and the schematic diagram is shown in **Figure 7a**. **Figure 7b** shows the $I_{ds}$ curves as a function of $V_g$ at various illumination power densities, exhibiting an obvious gate tunability of the photocurrent response of the Se nanosheet phototransistor. The device exhibits a pronounced photoresponse even at a very low illumination power down to 0.21 mW/cm$^2$, and the on-state photocurrent could reach up to 54 nA (**Figure S9a**). The enhancement of conduction under illumination confirms that Se nanosheet could be used for low-noise, high-sensitivity optoelectronic applications.

Photoresponsivity ($R_\lambda$), defined as $R_\lambda = I_{ph}/PS$, is calculated to better evaluate the performance of phototransistor, where $I_{ph}$ is the generated photocurrent, $P$ is the incident power, and $S$ is the effective illuminated area. $R_\lambda$ is estimated to be 263 A/W at an illumination power of 0.21 mW/cm$^2$ with $V_{ds}$=3V and $V_{bg}$=−80V. This value is four orders of magnitude higher than the previously reported crystalline Se films grown by physical vapor deposition (17 mA/W with a bias of −10 V) and among the highest values reported for 2D materials.[43-47] The responsivity is linearly proportional to the power of the illumination as shown in **Figure 7c**, suggesting that the photocurrent is mainly determined by the photoexcited carriers.[48] Time-dependent photoresponse of laser ON and OFF is also measured at room temperature (**Figure 7d**). The response speed is characterized by a typical rise time of 0.10 s and decay time 0.12 s for $V_{ds}$=3V and $V_{bg}$=0 V (taking 10−90% photocurrent change for the rise times and 90−10% for the fall times). **Figure S9 b** shows the photoresponse after 4 illumination cycles, the same level of photocurrent and noise with laser switch demonstrate a stable and repeatable photoresponse of our device.

**Conclusion**

In summary, high-quality 2D Se nanosheet were successfully synthesized by physical vapor deposition method. The as-synthesized Se nanosheet has a large lateral size up to 30 $\mu$m and the minimum thickness of 5 nm. The crystal structure of 2D Se nanosheet and its growth mechanism have been studied. Back-gated FETs and four-terminal devices based on Se nanosheet have been demonstrated, and the intrinsic carrier mobility is determined to be 0.26 cm$^2$ V$^{-1}$ s$^{-1}$ at 300K. The 2D Se nanosheet phototransistor exhibits an excellent photoresponsivity of 263 A/W, although the response time is slow. As a class of 2D materials formed by 1D van der Waals materials, Se nanosheet would have a great potential in electronic and optoelectronic applications.

**Experimental Methods**

**Growth of selenium nanosheets.** High purity Se power (Sigma-Aldrich, 99.99%) was placed at the center of heating zone in a multi-zone furnace, with a freshly cleaned Si (111) substrate located about 20~25 cm away from powders. During growth process, pure Ar gas was fed with a constant flow rate of 50 sccm. The tube pressure was maintained constant at 100 mbar. The whole reaction process was carried out under temperature of 210 ºC for the source and 100 ºC for the substrate, and this temperature was maintained for 60 min. After the reaction completed, black-colored needlelike materials coated on Si substrate could be clearly indentified.

**Raman and STEM measurements.** Raman measurement was performed using a HORIBA LabRAM HR800 Raman spectrometer. The system is equipped with a He-Ne excitation laser of 633 nm wavelength. The system was calibrated with the Raman peak of Si at 520 cm$^{-1}$ before measurement. The incident laser was polarized along [0001] direction of the selenium nanosheets, and illuminated perpendicularly to the nanosheet surface. The polarized laser is parallel to spiral atom chains and we denote this configuration as 0°. To avoid destroying sample, the laser power is less than 1 mW. The HAADF-STEM were performed with FEI Talos F200x equipped with a so-called probe corrector. This microscope was operated with an acceleration voltage of 200 kV.

**Device fabrication and characterization.** Selenium nanosheets were transferred onto 300nm SiO$_2$/Si substrates using scotch tape method. Electron beam lithography was used to pattern electrodes, followed by electron beam evaporation of 30nm Ni and 100nm Au as metal contacts. The channel length $L_{ch}$ was designed to be 4 μm while the channel width $W_{ch}$ = 10 μm. Four terminal field-effect devices are also fabricated to evaluate the intrinsic hole mobility of Se nanosheet, the channel length $L_{ch}$ between two voltage probes is 8.0 μm, and channel width $W_{ch}$ is 4.0 μm. The devices were measured with a probe station connected to semiconductor characterization system (4200SCS, Keithley) at room temperature. For photodetection, a 637nm laser source (S1FC637, Perot Benchtop) calibrated by an UV-enhanced silicon photodiode was used to provide power-tunable irradiation.

**FIGURES**

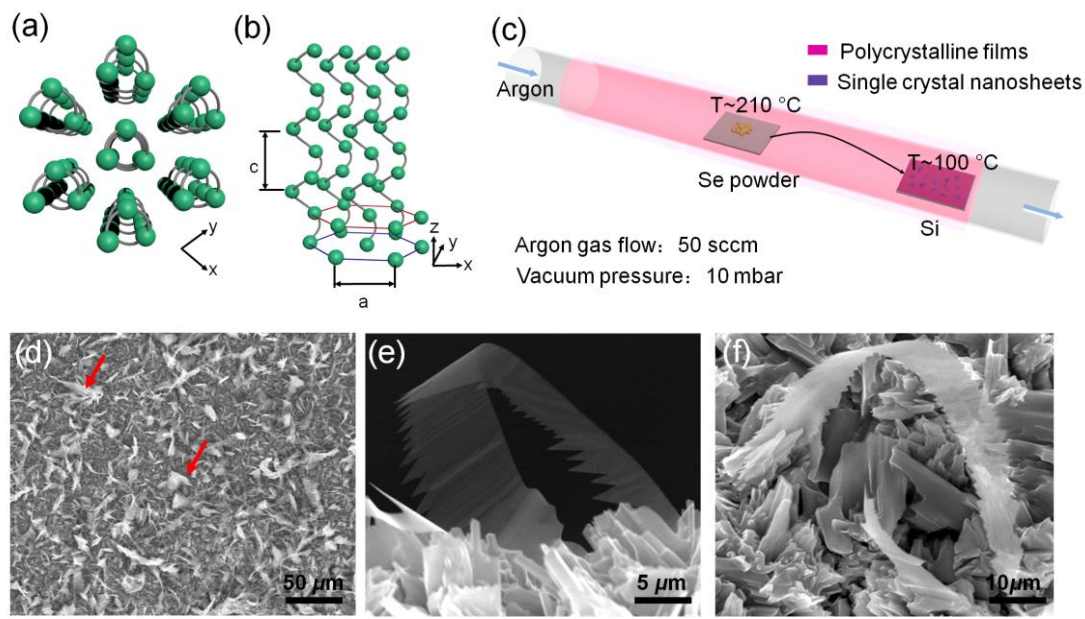

**Figure 1 | PVD-grown large-area Se nanosheets and material characterization.** (a-b) Atomic structure of selenium. (c) Schematic diagram of the PVD method. (d) Low-magnification SEM image of as-prepared selenium nanosheets on Si (111) substrate. (e) Enlarged view of typical Se nanosheets with saw-like structure and (f) feather-like twin structure.

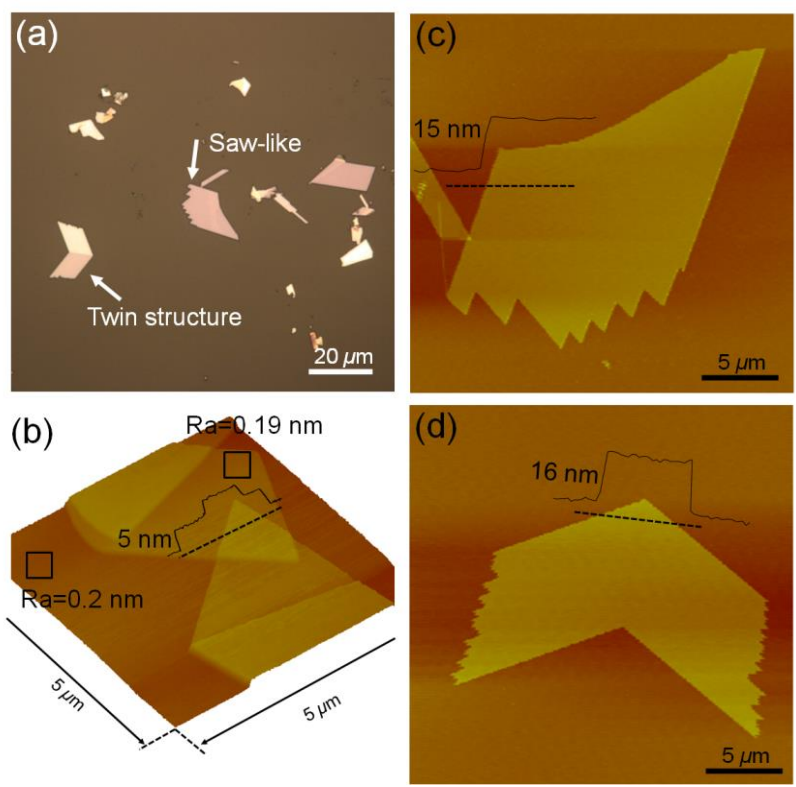

**Figure 2 | Optical and AFM characterization.** (a) Optical microscopy image of samples after transferred onto the SiO$_2$/Si substrate. (b) AFM topography image of Se nanosheet with thickness 5nm in 3D view, showing an atomically flat surface of Se nanosheet. (c-d) AFM height profile of saw-like and twin structure nanosheet.

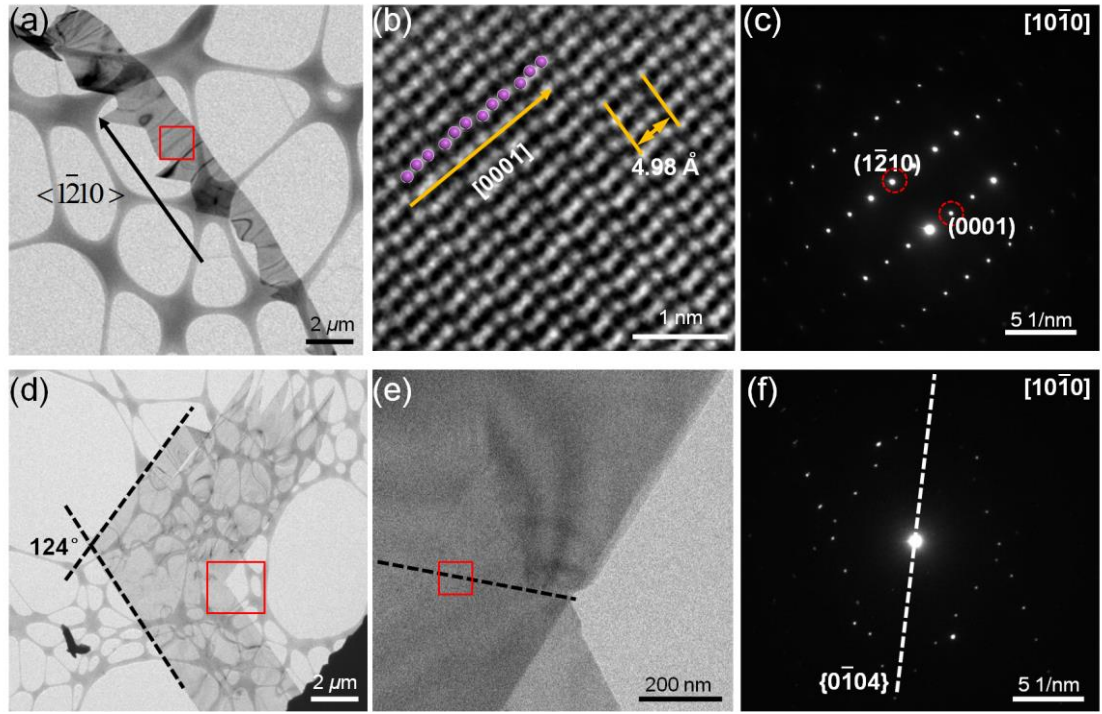

**Figure 3 | STEM characterization of Se nanosheet**. (a) Bright-field TEM image of Se nanosheets with saw-like structure. (b) HAADF-STEM image and (c) corresponding SAED pattern of Se nanosheets (boxed area in (a)). (d-e) Bright-field TEM image of feather-like Se nanosheets with different magnification. (f) SAED pattern obtained at the coherent crystal boundary (boxed area in (e)), common spots showing the mutual twin boundary of the $\{0\bar{1}04\}$ plane.

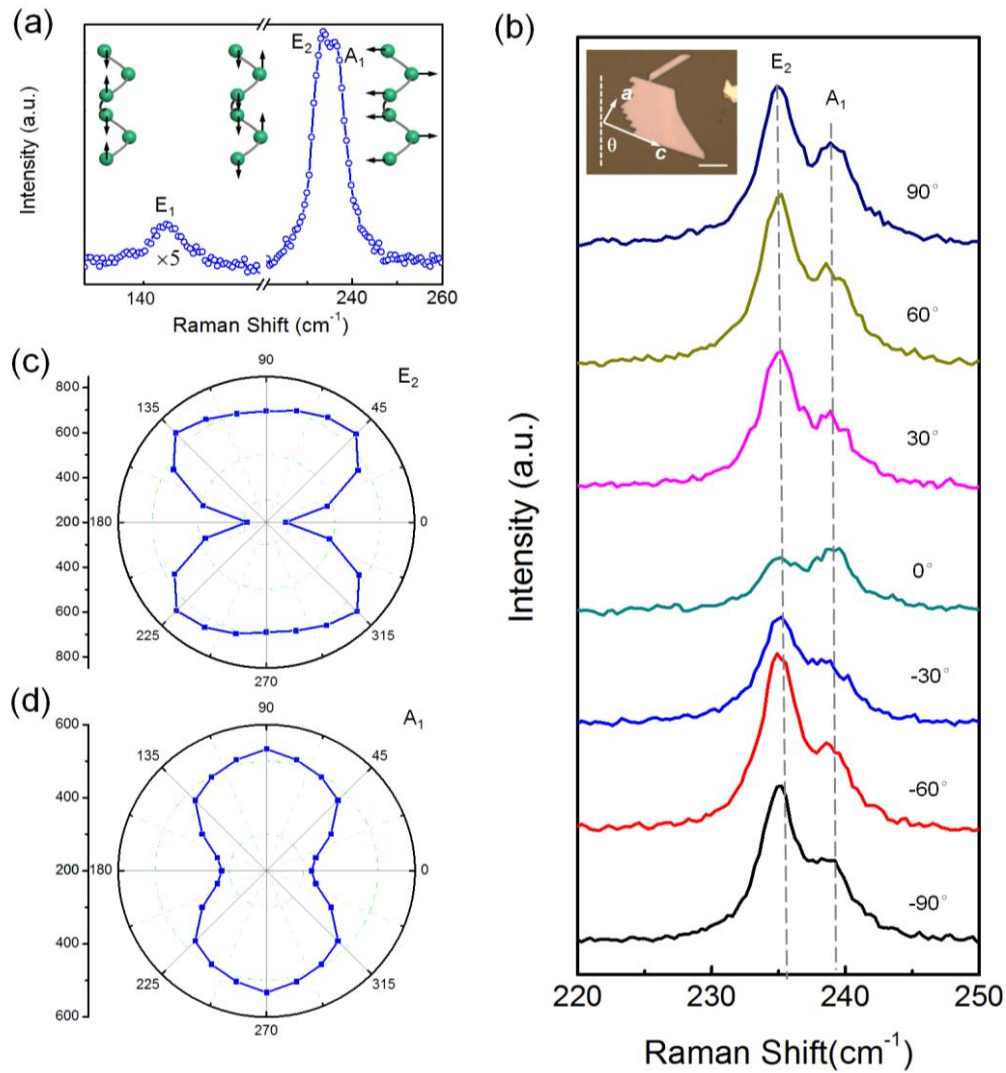

**Figure 4 | Angle-resolved Raman Spectra for few-layer Se nanosheets.** (a) Raman spectrum of Se nanosheet with thickness of 15 nm, inset shows the atomic vibrationally patterns of $E_1$, $E_2$, and $A_1$ phonon modes in selenium. (b) Raman spectra evolution with angles between crystal orientation and incident laser polarization. (c-d) Polar figures of Raman Intensity corresponding to $E_2$ and $A_1$ modes located at 233 and 237 cm$^{-1}$.

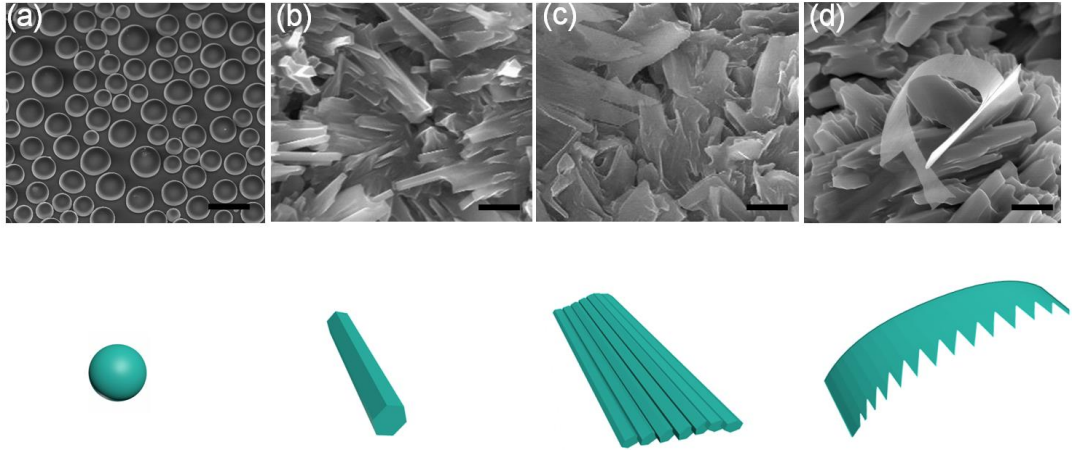

**Figure 5 | Growth mechanism of Se nanosheets.** (a)160°C. Scale bar is 20 μm. (b) 210°C, 5 min.(c) 210°C, 30 min. (d) 210°C, 60 min. Scale bar in (b-c) is 5 μm

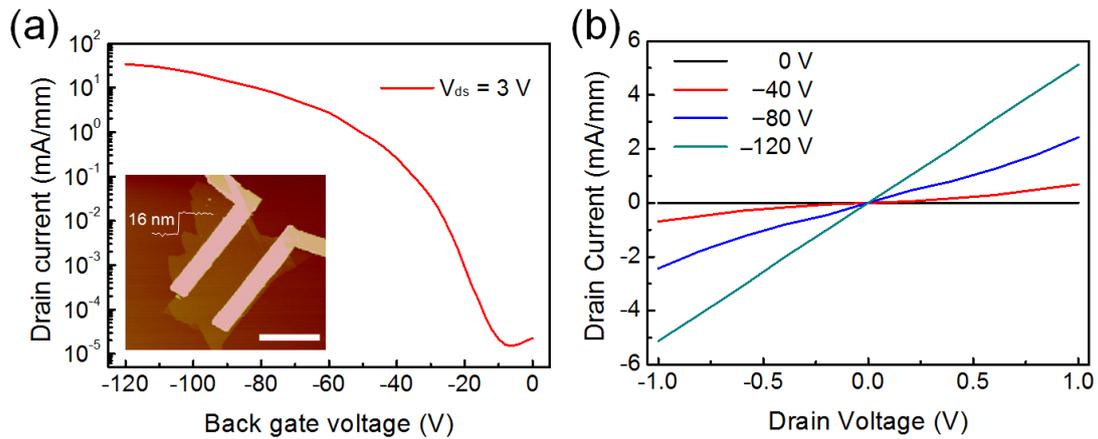

**Figure 6 | Se nanosheet FET device performance.** (a) Transfer characteristic of a typical Se nanosheet field effect transistor with thickness of 16 nm. Inset shows the AFM height profile of the device, scale bar is 5 μm. (b) Output characteristic of the same Se nanosheet transistor.

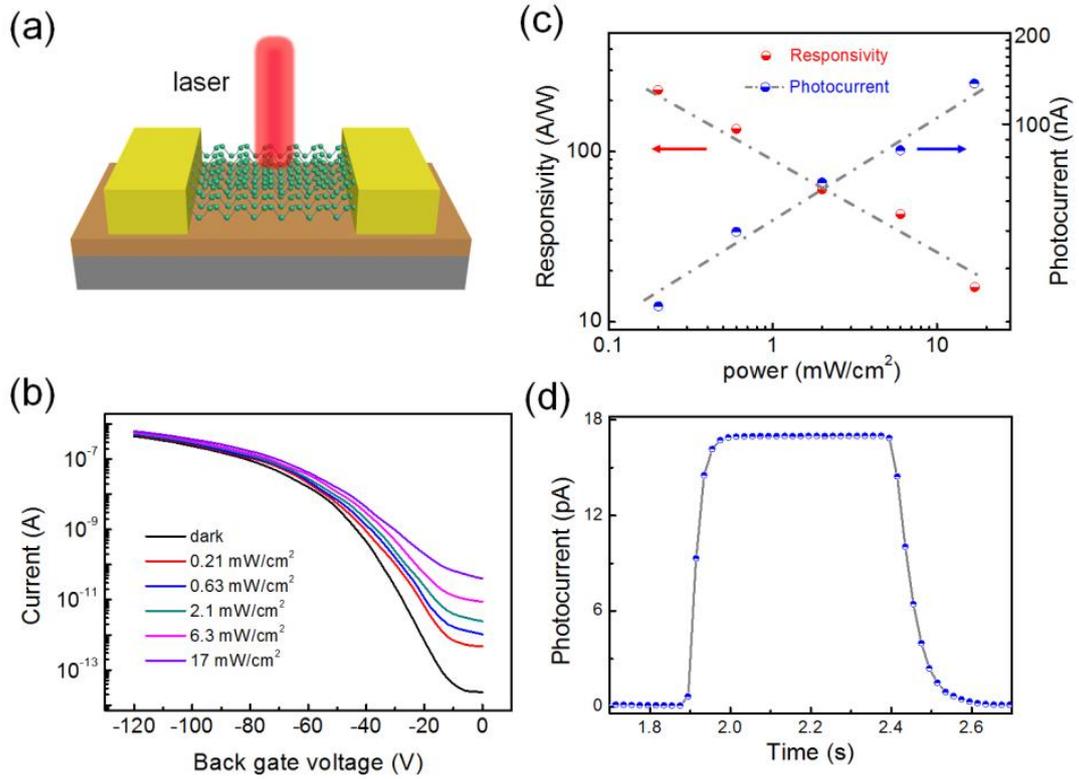

**Figure 7 | Photoelectrical properties of Se nanosheet phototransistor.** (a) Schematic diagram of a back-gated Se nanosheet phototransistor. (b)Transfer curves of Se nanosheet phototransistor measured under various laser irradiation powers at $V_{ds}$ =3 V. (c) Photocurrent ($I_{ph}$) and responsivity ($R_\lambda$) as a function of laser illumination power measured at $V_{ds}$ = 3 V and $V_g$ = −80 V. The dotted line for $I_{ph}$ is well match with the data using the function $I_{ph} = P^\beta$, where $P$ is the effective power of the illumination on the device and $\beta$ is a constant. (d) Typical rise or decay characteristics of the photocurrent with the laser illumination switched on or off.

References

1. Kasap, S.; Frey, J. B.; Belev, G.; Tousignant, O.; Mani, H.; Laperriere, L.; Reznik, A.; Rowlands, J. A. Amorphous selenium and its alloys from early xeroradiography to high resolution X-ray image detectors and ultrasensitive imaging tubes. *Phys. Status Solidi B.* **2009**, 246, 1794-1805.
2. Suzuki, Y.; Yamaguchi, H.; Oonuki, K.; Okamura, Y.; Okano, K. Amorphous selenium photodetector driven by diamond cold cathode. *IEEE Electron Device Lett.* **2003**, 24, 16-18.
3. Oonuki, K.; Suzuki, Y.; Yamaguchi, H.; Okano, K.; Okamura, Y. Diode structure amorphous selenium photodetector with nitrogen (N)-doped diamond cold cathode. *J. Vac. Sci. Technol., B: Microelectron. Nanometer Struct. Process., Meas., Phenom.* **2003**, 21, 1586-1588.
4. Wang, K.; Chen, F.; Belev, G.; Kasap, S.; Karim, K. S. Lateral metal-semiconductor-metal photodetectors based on amorphous selenium. *Appl. Phys. Lett.* **2009**, 95, 013505.
5. Abbaszadeh, S.; Allec, N.; Wang, K.; Karim, K. S. Low dark-current lateral amorphous-selenium metal–semiconductor–metal photodetector. *IEEE Electron Device Lett.* **2011**, 32, 1263-1265.
6. Royer, D.; Dieulesaint, E. Elastic and piezoelectric constants of trigonal selenium and tellurium crystals. *J. Appl. Phys.* **1979**, 50, 4042-4045.
7. Lee, T. I.; Lee, S.; Lee, E.; Sohn, S.; Lee, Y.; Lee, S.; Moon, G.; Kim, D.; Kim, Y. S.; Myoung, J. M. High-Power Density Piezoelectric Energy Harvesting Using Radially Strained Ultrathin Trigonal Tellurium Nanowire Assembly. *Adv. Mater.* **2013,** 25, 2920-2925.
8. He, Z.; Yang, Y.; Liu, J.-W.; Yu, S.-H. Emerging tellurium nanostructures: controllable synthesis and their applications. *Chem. Soc. Rev.* **2017**,46, 2732-2753.
9. Abad, B.; Rull-Bravo, M.; Hodson, S. L.; Xu, X.; Martin-Gonzalez, M. Thermoelectric properties of electrodeposited tellurium films and the sodium lignosulfonate effect. *Electrochim. Acta*. **2015**, 169, 37-45.
10. Sridharan, K.; Ollakkan, M. S.; Philip, R.; Park, T. J. Non-hydrothermal synthesis and optical limiting properties of one-dimensional Se/C, Te/C and Se–Te/C core–shell nanostructures. *Carbon* **2013**, 63, 263-273.
11. Wang, R.; Su, X.; Bulla, D.; Wang, T.; Gai, X.; Yang, Z.; Madden, S.; Luther-Davies, B. In Identifying the best chalcogenide glass compositions for the application in mid-infrared waveguides. *Proc. SPIE*. **2015**, *DOI. 10.1117/12.2074815.*
12. Qian, J.; Jiang, K. J.; Huang, J. H.; Liu, Q. S.; Yang, L. M.; Song, Y. A Selenium-Based Cathode for a High-Voltage Tandem Photoelectrochemical Solar Cell. *Angew. Chem., Int. Ed.* **2012**, 51, 10351-10354.
13. Nguyen, D.-C.; Tanaka, S.; Nishino, H.; Manabe, K.; Ito, S. 3-D solar cells by electrochemical-deposited Se layer as extremely-thin absorber and hole conducting layer on nanocrystalline TiO2 electrode. *Nanoscale Res. Lett.* **2013**, 8, 8.
14. Zheng, L.; Hu, K.; Teng, F.; Fang, X. Novel UV–Visible Photodetector in Photovoltaic Mode with Fast Response and Ultrahigh Photosensitivity Employing Se/TiO2 Nanotubes Heterojunction. *Small.***2016**,13,1602448.
15. Mondal, K.; Roy, P.; Srivastava, S. K. Facile biomolecule-assisted hydrothermal synthesis of trigonal selenium microrods. *Cryst. Growth Des.* **2008**, 8, 1580-1584.
16. Ren, L.; Zhang, H.; Tan, P.; Chen, Y.; Zhang, Z.; Chang, Y.; Xu, J.; Yang, F.; Yu, D. Hexagonal selenium nanowires synthesized via vapor-phase growth. *J. Phys. Chem. B.* **2004**, 108, 4627-4630.

17. Luo, L.-B.; Yang, X.-B.; Liang, F.-X.; Jie, J.-S.; Li, Q.; Zhu, Z.-F.; Wu, C.-Y.; Yu, Y.-Q.; Wang, L. Transparent and flexible selenium nanobelt-based visible light photodetector. *CrystEngComm.* **2012**, 14, 1942-1947.
18. Reza, S.; Bosman, G.; Islam, M. S.; Kamins, T. I.; Sharma, S.; Williams, R. S. Noise in silicon nanowires. *IEEE Trans. Nanotechnol.* **2006**, 5, 523-529.
19. Bid, A.; Bora, A.; Raychaudhuri, A. 1/f noise in nanowires. *Nanotechnology.* **2005**, 17, 152.
20. Wang, Y.; Qiu, G.; Wang, Q.; Liu, Y.; Du, Y.; Wang, R.; Goddard III, W. A.; Kim, M. J.; Ye, P. D.; Wu, W. Large-area solution-grown 2D tellurene for air-stable, high-performance field-effect transistors. *arXiv preprint arXiv.* **2017**.*1704.06202*.
21. Du, Y.; Qiu, G.; Wang, Y.; Si, M.; Xu, X.; Wu, W.; Ye, P. D. 1D van der Waals Material Tellurium: Raman Spectroscopy under Strain and Magneto-transport. *arXiv preprint arXiv.***2017**.*1704.07020*.
22. Wang, Q.; Safdar, M.; Xu, K.; Mirza, M.; Wang, Z.; He, J. Van der Waals epitaxy and photoresponse of hexagonal tellurium nanoplates on flexible mica sheets. *Acs Nano.* **2014**, 8, 7497-7505.
23. Jiang, Z.-Y.; Xie, Z.-X.; Xie, S.-Y.; Zhang, X.-H.; Huang, R.-B.; Zheng, L.-S. High purity trigonal selenium nanorods growth via laser ablation under controlled temperature. *Chem. Phys. Lett.* **2003**, 368, 425-429.
24. Shenasa, M.; Sainkar, S.; Lichtman, D. XPS study of some selected selenium compounds. *J. Electron Spectrosc. Relat. Phenom.* **1986**, 40, 329-337.
25. Martin, R. M.; Lucovsky, G.; Helliwell, K. Intermolecular bonding and lattice dynamics of Se and Te. *Phys. Rev. B.* **1976**, 13, 1383.
26. Lucovsky, G.; Mooradian, A.; Taylor, W.; Wright, G.; Keezer, R. Identification of the fundamental vibrational modes of trigonal, α-monoclinic and amorphous selenium. *Solid State Commun.* **1967**, 5, 113-117.
27. Serra, A.; Rossi, M.; Buccolieri, A.; Manno, D.; Rossi, M.; Mariani, C.; Terranova, M. L. Solid-to-solid phase transformations of nanostructured selenium-tin thin films induced by thermal annealing in oxygen atmosphere. *AIP Conf. Proc.* 2**014**, 31-39.
28. Imura, S.; Kikuchi, K.; Miyakawa, K.; Ohtake, H.; Kubota, M.; Nakada, T.; Okino, T.; Hirose, Y.; Kato, Y.; Teranishi, N. High sensitivity image sensor overlaid with thin-film crystalline-selenium-based heterojunction photodiode . *IEEE Electron Device Lett.* **2016**, 63, 86-91.
29. Wang, Q.; Li, G.-D.; Liu, Y.-L.; Xu, S.; Wang, K.-J.; Chen, J.-S. Fabrication and growth mechanism of selenium and tellurium nanobelts through a vacuum vapor deposition route. *J. Phys. Chem. C* **2007**. 111, 12926-12932.
30. Hawley, C. J.; Beatty, B. R.; Chen, G.; Spanier, J. E. Shape-controlled vapor-transport growth of tellurium nanowires. *Cryst. Growth Des.* **2012**, 12, 2789-2793.
31. Pan, Z. W.; Dai, Z. R.; Wang, Z. L. Nanobelts of semiconducting oxides. *Science.* **2001**, 291, 1947-1949.
32. Hu, Z.; Duan, X.; Gao, M.; Chen, Q.; Peng, L.-M. ZnSe nanobelts and nanowires synthesized by a closed space vapor transport technique. *J. Phys. Chem. C.* **2007**, 111, 2987-2991.
33. Gamalski, A. D.; Voorhees, P. W.; Ducati, C.; Sharma, R.; Hofmann, S. Twin plane re-entrant mechanism for catalytic nanowire growth. *Nano Lett.* **2014**, 14, 1288-1292.
34. Soo, M. T.; Zheng, K.; Gao, Q.; Tan, H. H.; Jagadish, C.; Zou, J. Mirror-twin induced bicrystalline InAs nanoleaves. *Nano Res.* **2016**, 9, 766-773.
35. Sheu, J.; Su, Y.-K.; Chi, G.-C.; Koh, P.; Jou, M.; Chang, C.; Liu, C.; Hung, W. High-transparency

Ni/Au ohmic contact to p-type GaN. *Appl. Phys. Lett.* **1999**, 74, 2340-2342.

36. Belev, G.; Tonchev, D.; Fogal, B.; Allen, C.; Kasap, S. Effects of oxygen and chlorine on charge transport in vacuum deposited pure a-Se films. *J. Phys. Chem. Solids.* **2007**, 68, 972-977.

37. Iqbal, M. W.; Iqbal, M. Z.; Khan, M. F.; Shehzad, M. A.; Seo, Y.; Park, J. H.; Hwang, C.; Eom, J. High-mobility and air-stable single-layer WS2 field-effect transistors sandwiched between chemical vapor deposition-grown hexagonal BN films. *Sci. Rep.* **2015**, 5, 10699.

38. Radisavljevic, B.; Radenovic, A.; Brivio, J.; Giacometti, i. V.; Kis, A. Single-layer MoS2 transistors. *Nat. Nanotechnol.* **2011**, 6, 147-150.

39. Liu, H.; Du, Y.; Deng, Y.; Ye, P.D.. Semiconducting black phosphorus: synthesis, transport properties and electronic applications. *Chem. Soc. Rev.* **2015**, 44, 2732-2743.

40. Sharma, A. K.; Singh, B. Electrical conductivity measurements of evaporated selenium films in vacuum. *Proc. Indian Natn. Sci. Acad.* **1980**, 362-368.

41. Spear, W. The hole mobility in selenium. *Proc. Phys. Soc.* **1960**, 76, 826.

42. Plessner, K. Conductivity, Hall Effect and Thermo-electric Power of Selenium Single Crystals. *Proc. Phys. Soc.* **1951**, 64, 671.

43. Zhai, T.; Fang, X.; Liao, M.; Xu, X.; Li, L.; Liu, B.; Koide, Y.; Ma, Y.; Yao, J.; Bando, Y. Fabrication of high-quality In2Se3 nanowire arrays toward high-performance visible-light photodetectors. *Acs Nano,* **2010**, 4, 1596-1602.

44. Tao, X.; Gu, Y. Crystalline–Crystalline Phase Transformation in Two-Dimensional In2Se3 Thin Layers. *Nano lett.* **2013**, 13, 3501-3505.

45. Xia, J.; Zhu, D.; Wang, L.; Huang, B.; Huang, X.; Meng, X. M. Large-Scale growth of two-dimensional SnS2 crystals driven by screw dislocations and application to photodetectors. *Adv. Funct. Mater.* **2015**, 25, 4255-4261.

46. Zhou, X.; Zhang, Q.; Gan, L.; Li, H.; Zhai, T. Large-Size Growth of Ultrathin SnS2 Nanosheets and High Performance for Phototransistors. *Adv. Funct. Mater.* **2016**, 26, 4405-4413.

47. Yin, Z.; Li, H.; Li, H.; Jiang, L.; Shi, Y.; Sun, Y.; Lu, G.; Zhang, Q.; Chen, X.; Zhang, H. Single-layer MoS2 phototransistors. *Acs Nano.* **2011**, 6, 74-80.

48. Li, X.; Cui, F.; Feng, Q.; Wang, G.; Xu, X.; Wu, J.; Mao, N.; Liang, X.; Zhang, Z.; Zhang, J. Controlled growth of large-area anisotropic ReS 2 atomic layer and its photodetector application. *Nanoscale.* **2016**, 8, 18956-18962.

ACKNOWLEDGEMENTS

The authors would like to thank Chao Wang at Harbin Institute of Technology for the discussion in Raman measurements. J.K. Qin is supported by Chinese Scholarship Council. The work is in part supported by AFOSR/NSF EFRI 2DARE program, ARO and SRC.

AUTHOR CONTRIBUTIONS

P.D.Y. conceived the idea and supervised the experiments. J.K.Q performed the growth experiments and analyzed the experimental data. J.K.Q. and G.Q., performed device fabrication and analyzed the experimental data. C.Y.X and W.Z. Wu analyzed the growth mechanism. H.Z. performed the SEM measurement. J.J. performed the STEM measurement. L.M.Y and X.F.X. performed Raman measurement. A. R. C and D. Z performed and supervised the XPS analysis, H.Y.W. analyzed and supervised STEM experiment. J.K.Q., G.Q., and P.D.Y. co-wrote the manuscript.

COMPETING FINANCIAL INTERESTS STATEMENT

The authors declare no competing financial interests.

Supporting Information Available*:* The schematic diagram of transfer procedure, XPS, EDX and EELS spectrum of Se nanosheets, SEM image of Se microspheres and nanowires, the air stability of Se nanosheet, four-terminal devices and related measurement, and photoelectronic properties of Se nanosheet. This material is available free of charge *via* the Internet at *http://pubs.acs.org*.

Supplementary Information for:

# Controlled Growth of Large-Size 2D Selenium Nanosheet and Its Electronic and Optoelectronic Applications

Jing-Kai Qin[1,3,†], Gang Qiu[1,†], Jie Jian[2], Hong Zhou[1], Ling-Ming Yang[1], Adam R. Charnas[1], Dmitry Zemlyanov[1], Cheng-Yan Xu[3], Xian-Fan Xu[4], Wen-Zhuo Wu[5], Hai-Yan Wang[2,1], Peide D. Ye[1,*]

[†] These authors contributed equally to this work.

[1.] School of Electrical and Computer Engineering, Purdue University, West Lafayette, IN 47907, USA

[2.] School of Materials Science and Engineering, Purdue University, West Lafayette, In 47907, USA

[3.] School of Materials Science and Engineering, Harbin Institute of Technology, Harbin 150001, Peoples' Republic of China.

[4.] School of Mechanical Engineering, Purdue University, West Lafayette, In 47907, USA

[5.] School of Industry Engineering, Purdue University, West Lafayette, In 47907, USA

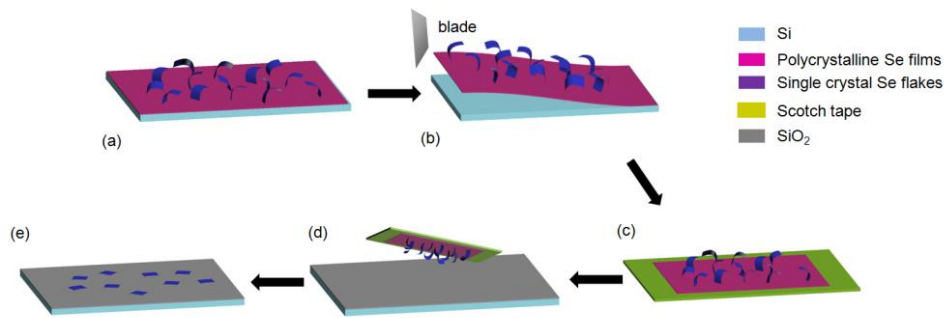

**Supplementary Figure S1| Schematic diagram of the transfer procedure.** 1) use a shape blade to detach the polycrystalline films and silicon substrate; 2) transfer flakes/films to the scotch tape; 3) put the tape face-down on SiO$_2$/Si substrate and press it slightly with a rubber block; 4) peel off the tape and Se flakes would be left on substrate surface.

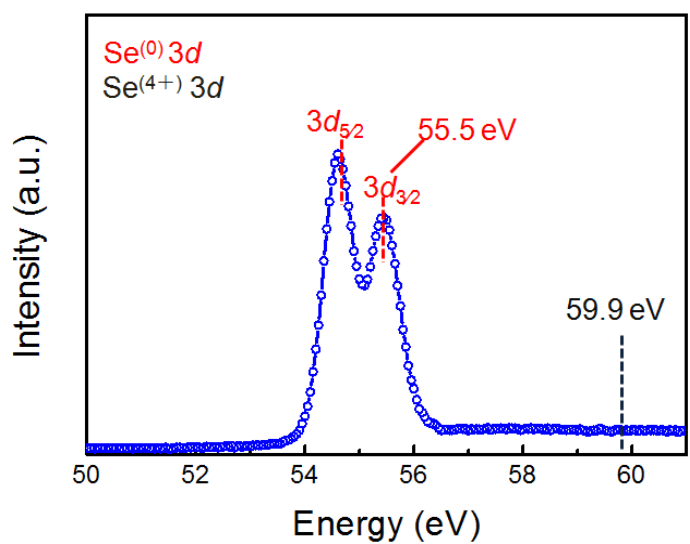

**Supplementary Figure S2 | XPS spectrum of Se nanosheet.** The strong peaks located at 55.5 and 54.3 eV correspond to Se $3d_{3/2}$ and $3d_{5/2}$ binding energy for $Se^{(0)}$, and Se 3d oxidized peak located 59.9 eV cannot detected, indicating the Se nanosheets are elemental crystals without oxidation.

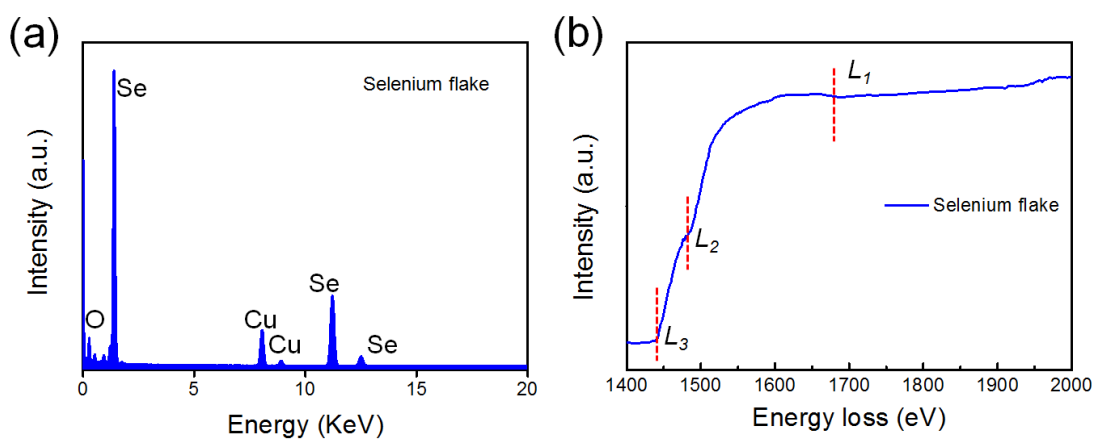

**Supplementary Figure S3 | Element analysis of few-layer Se nanosheet.** (a) EDX spectrum of Se nanosheet, peaks of Cu are from the copper grid. (b) EELS spectrum of Se nanosheet, the selenium L electron energy loss (EEL) edges are detected at 1435 eV ($L_3$), 1477 eV ($L_2$) and 1659 eV ($L_1$), respectively, indicating the nanosheet is

highly crystalline selenium.

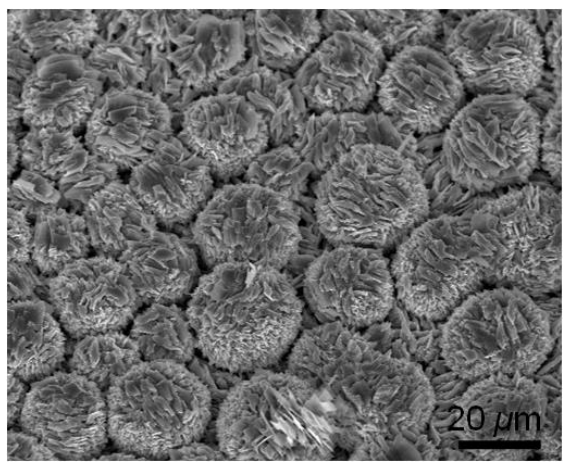

**Supplementary Figure S4 | SEM image of the Se polycrystalline nanospheres.** Short nanorods and thick plates protruding out of the surface of amorphous nanospheres start to appear.

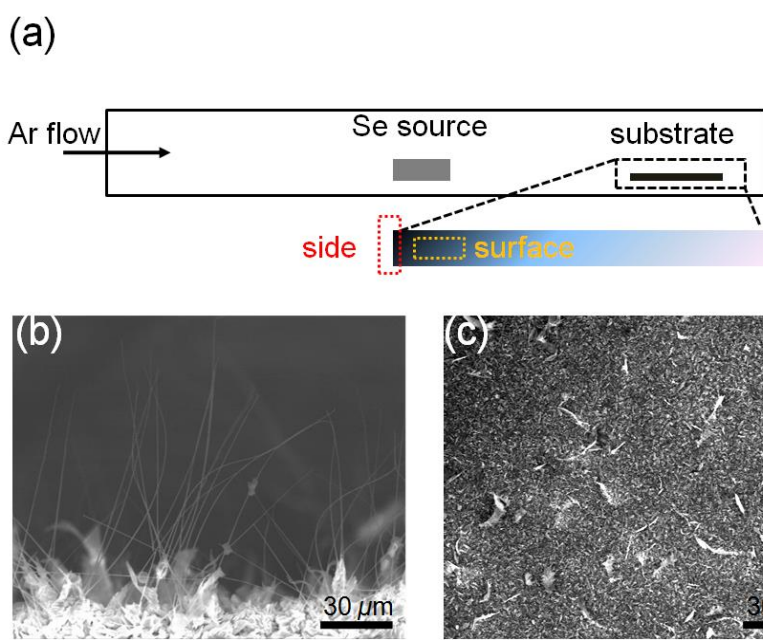

**Supplementary Figure S5 | The effect of substrate position on the product morphology.** (a) Schematic layout of the PVD system. (b) SEM image of the Se nanowires grown on the side of substrate. (c) SEM image of the Se flakes grown on

the surface of substrate.

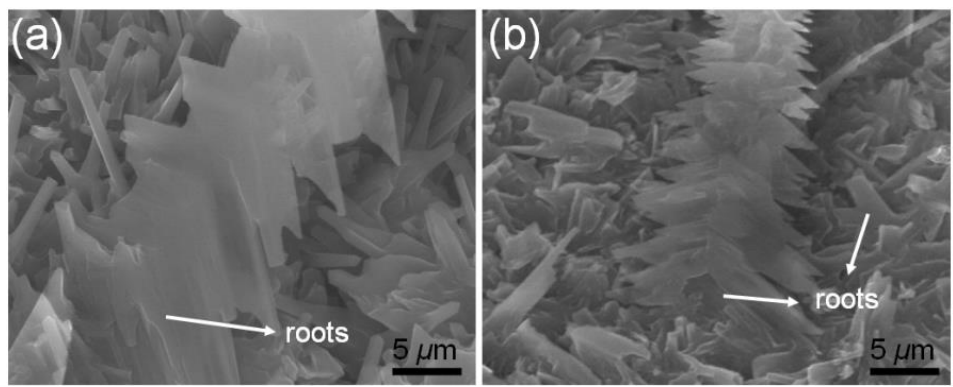

**Supplementary Figure S6 | The growth roots for selenene twinned structure.** (a,b) SEM image of feather-like nanoflakes with twin structure, crossed nanorod could be served as root for growth.

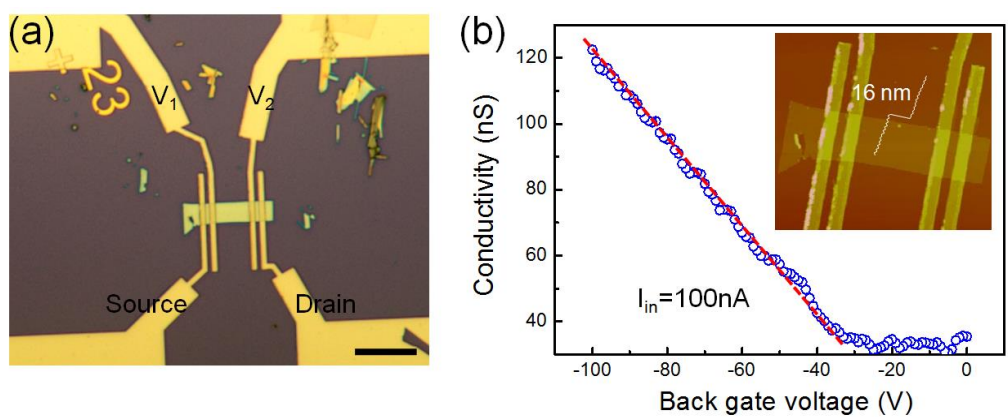

**Supplementary Figure S7 | Electrical transport measurement in four-terminal back-gated selenium nanosheet device.** (a) Optical microscopy image of the device, fixed channel current $I_{in}$ of 100 nA is applied on Source and Drain probes, voltage drop $V_d$ across the channel between probes $V_1$ and $V_2$ is recorded. The scale bar is 10 μm. (b) Conductance as a function of back-gate voltage measured at 300 K. The inset shows the AFM image of the four-terminal device.

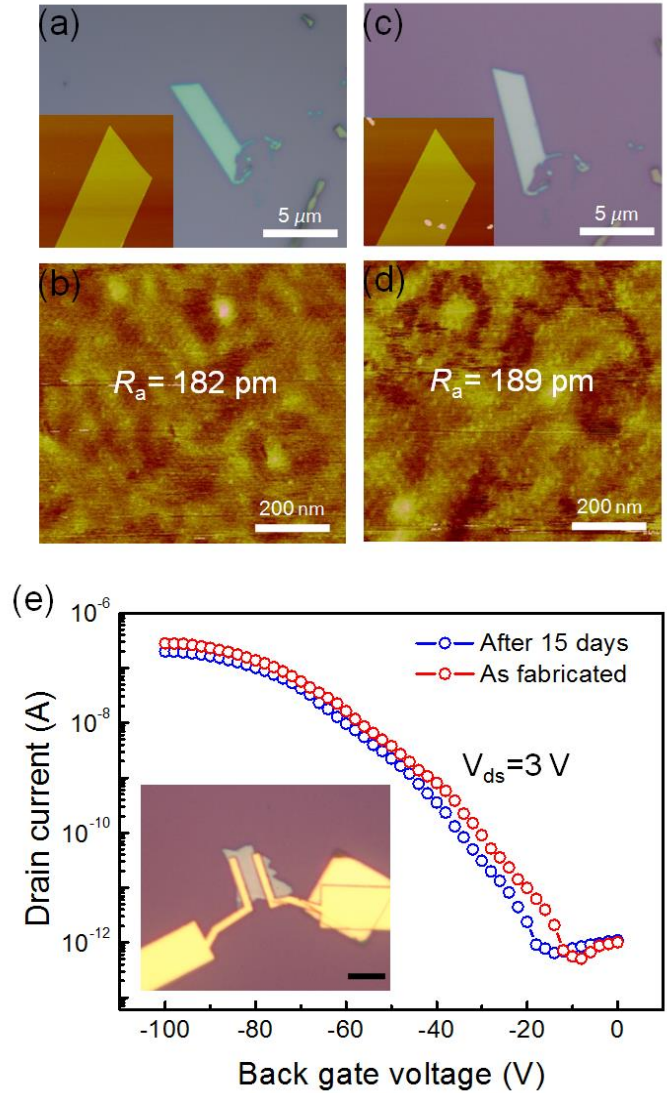

**Supplementary Figure S8 | Air stability of Se nanosheet** (a, b) Surface topography evolution of selenium flakes after exposing in air condition for 15 days. Inset: the AFM height profile. (c, d) Corresponding surface topography profile. (e) Transfer characteristic of a typical Se nanosheet field effect transistor as fabricated and after 15 days storage in air.

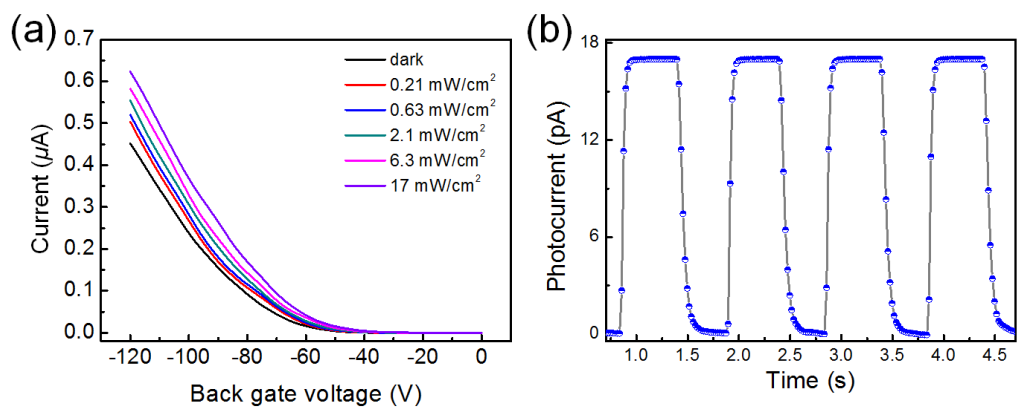

**Supplementary Figure S9 | Photoelectrical performance of selenene phototransistor.** (a) Transfer curves of selenene phototransistor measured under various laser illumination powers at $V_{ds}$ =3 V. (b) Time-resolved photoresponse of the selenene phototransistor recorded by alternative switching on and off the laser illumination.

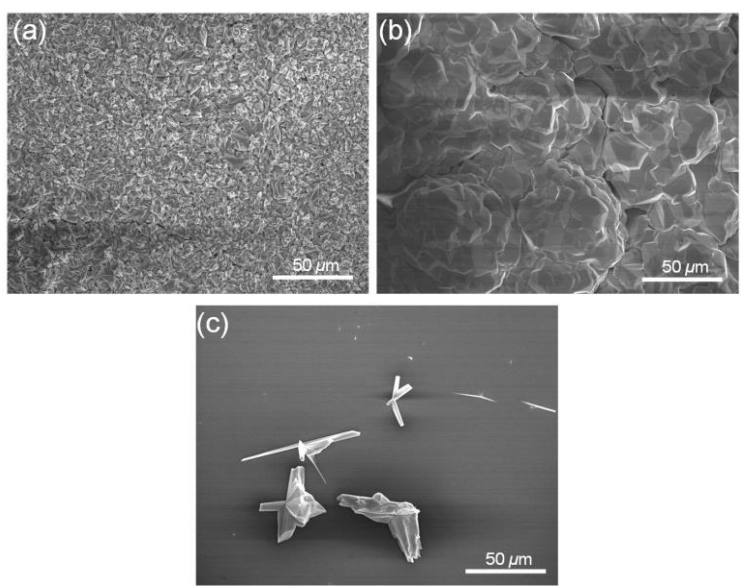

**Supplementary Figure S10 | SEM images of products grown on different substrates.** (a) Sapphire (b) Mica (c) $SiO_2$